\documentclass[12pt]{article}
\usepackage{amssymb}
\usepackage{amsmath}
\usepackage{graphicx}
\usepackage{cite}

\begin{document}





\begin{center}

{\large Perturbativity in the seesaw mechanism
{\bf } 
}

\vskip 0.5cm

Takehiko Asaka$^a$, Takanao Tsuyuki$^b$\\

\vskip 0.4cm

{\it $^a$Department of Physics, Niigata University, 950-2181 Niigata, Japan}\\
{ \it$^b$Graduate School of Science and Technology, Niigata University, Niigata 950-2181, Japan}

\vskip 0.2cm
(Septemter 9, 2015)

\begin{abstract}
We consider the Standard Model extended by right-handed neutrinos to explain massive neutrinos through the seesaw mechanism. The new fermion can be observed when it has a sufficiently small mass and large mixings to left-handed neutrinos. If such a particle is the lightest right-handed neutrino, its contribution to the mass matrix of active neutrinos needs to be canceled by that of a heavier one. Yukawa couplings of the heavier one are then larger than those of the lightest one. We show that the perturbativity condition gives a severe upper bound on the mixing of the lightest right-handed neutrino, depending on the masses of heavier ones. Models of high energy phenomena, such as leptogenesis, can be constrained by low energy experiments.
\end{abstract}

\end{center}

\vskip .5cm

\section{Introduction}
The masses of active neutrinos are the most prominent evidence for physics beyond the Standard Model. The simplest and sufficient extension is to introduce right-handed neutrinos (the seesaw mechanism \cite{Minkowski:1977sc}). They can also explain the baryon asymmetry of the universe \cite{Fukugita:1986hr}. Right-handed neutrinos have been searched for a long time at various experiments and observations (for compilations, see Refs. \cite{Smirnov:2006bu,Atre:2009rg,Deppisch:2015qwa}). Their detectability is determined by two parameters: masses and mixings. 

In this letter, we derive upper and lower bounds on the mixing of the lightest right-handed neutrino. The lower bound comes from the seesaw relation, and the upper one is from the perturbativity. To obtain the upper bound, we take into account the cancellation between the contributions of right-handed neutrinos to the mass matrix of active neutrinos. We also consider the renormalization group evolution (RGE) of Yukawa couplings. The new upper bound is stronger than the existing bounds by orders of magnitude for many cases. This bound is crucial for future collider searches \cite{Adams:2013qkq,Blondel:2014bra,Kobach:2014hea,Antusch:2015mia,Banerjee:2015gca,Alekhin:2015byh,Asaka:2015oia}. 

The perturbativity of the neutrino Yukawa couplings without RGE was discussed in Ref. \cite{Casas:2010wm}. The RGE effect was partially studied in the context of grand unification \cite{Tsuyuki:2014xja}. Note that stronger bounds may be derived by the vacuum stability condition \cite{EliasMiro:2011aa,Rodejohann:2012px}, but it largely depends on the top Yukawa coupling and the details of the scalar sector. Here we focus on the perturbativity to be conservative. Our results do not change much even if we assume the stability. 

\section{Seesaw mechanism}
We consider the Standard Model extended by 
right-handed neutrinos $\nu_R$. The neutrino masses are induced by the Lagrangian
\begin{eqnarray}
  \label{eq:LAG}
  {\cal L}_{\rm mass} &=& - F_{\alpha I} \overline L_\alpha \Phi \nu_{R I}  - \frac{M_I}{2} \overline{\nu}_{RI} \nu_{R I}^c + h.c.
\end{eqnarray}
$F_{\alpha I}\,(\alpha=e,\mu,\tau;\, I=1, \dots, {\cal N})$ are Yukawa couplings of neutrinos and Higgs doublet $\Phi$. We assume $|F_{\alpha I}| \langle
\Phi \rangle/M_I \ll 1\; (\langle \Phi\rangle=174\,{\rm GeV})$ to realize the seesaw mechanism. 
 The mass eigenstates of neutrinos are three active neutrinos $\nu_j$ ($j =
1,2,3$) with masses $m_j$ and ${\cal N}$ heavy neutral leptons (HNLs)
$N_I$ with masses $M_I$. The labeling of HNLs is defined as
$M_1 \le M_2 \le  \dots \le M_{\cal N}$. The masses and mixings are related by the seesaw relation
\begin{align} 
   U D_\nu U^T=-\Theta D_N \Theta^T, \label{Eseesaw}
\end{align}
where $U$ denotes the active neutrino mixing matrix, $\Theta_{\alpha I}=F_{\alpha I} \langle
\Phi \rangle/M_I$ is the mixing element of HNL $N_I$, $D_\nu\equiv {\rm diag}(m_1, m_2, m_3)$ and $D_N\equiv {\rm diag}(M_1, \dots, M_{\cal N})$. The relation (\ref{Eseesaw}) is modified by radiative corrections  \cite{Grimus:2002nk,AristizabalSierra:2011mn}. These corrections do not change our final results since they give subleading contributions.

The Yukawa couplings which satisfy Eq. (\ref{Eseesaw}) are convenient to be parametrized as \cite{Casas:2001sr}
\begin{align}
F=\frac{i}{\langle\Phi\rangle}UD_\nu^{1/2}\Omega D_N^{1/2}, \label{eq:casas}
\end{align}
where $\Omega$ is a $3\times \cal N$ complex matrix which is not determined by neutrino oscillation experiments. It satisfies $\Omega\Omega^T=1$ for ${\cal N}\ge 3$ and $\Omega\Omega^T=$ diag(0,1,1) or diag(1,1,0) for ${\cal N}=2$ with the normal hierarchy (NH) or inverted hierarchy (IH). The mixing of the lightest HNL is expressed as
\begin{align}
|\Theta_1|^2\equiv\sum_\alpha |\Theta_{\alpha 1}|^2=\frac{1}{M_1}\sum_{j=1}^{3}m_j |\Omega_{j1}|^2. \label{Eth1}
\end{align}

We derive a lower bound on the mixing for ${\cal N}=2,\, 3$ cases,
\begin{align}
M_1|\Theta_1|^2
&=m_l\left(1-\sum_{j=1}^3(\Omega_{j1}^{r2}-\Omega_{j1}^{i2})\right)+\sum_{j=1}^3 m_j(\Omega_{j1}^{r2}+\Omega_{j1}^{i2})\notag\\
&=m_l+\sum_{j=1}^3 \left\{(m_j-m_l)\Omega_{j1}^{r2}+(m_j+m_l)\Omega_{j1}^{i2}\right\}\notag\\ 
&\geq m_l, \label{eq:low}
\end{align}
where $\Omega^r$ and $\Omega^i$ are real and imaginary parts of $\Omega$. $m_l$ is $m_2\, (m_1)$ for ${\cal N} =2$ with the NH (IH) and the lightest neutrino mass for ${\cal N}=3$.  In the first equality, we have used $(\Omega^T\Omega)_{11}=1$, which holds for ${\cal N}=2,\,3$. The last equality holds when $\Omega_{j1}=\delta_{jl}$. For ${\cal N}\ge 4$, there is no lower bound since we can choose $\Omega_{j1}$=0, which means that $N_1$ decouples from the seesaw mechanism. If the lightest neutrino is massless, $N_1$ can also decouple for ${\cal N}=3$.

\section{Perturbativity}
From Eq. (\ref{Eth1}), we can see that the mixing $|\Theta_1|^2$ can be much larger than $m_j/M_1$ if $|\Omega_{j1}|$ is much larger than 1. Actually, such $\Omega$ is needed to discover the HNL by collider experiments. For example, the bound from $Z$ boson decay is $|\Theta_1|^2\lesssim 10^{-5}$ for $M_1=10$ GeV \cite{Abreu:1996pa}, and then $M_1|\Theta_1|^2\sim 100\,{\rm keV}\gg m_j$  for $|\Theta_1|^2 \sim 10^{-5}$. In this case, the contribution of $N_1$ in the right-hand side of Eq. (\ref{Eseesaw}) need to be mostly canceled by those of heavier HNLs ($\Omega$ cannot be diagonal).

The mixing, however, cannot be arbitrarily large, since Yukawa couplings are constrained by perturbativity as discussed below. The diagonal components of the seesaw relation (\ref{Eseesaw}) are
\begin{align}
\sum_{j}m_j U_{\alpha j}^2=-\sum_{I}M_I\Theta_{\alpha I}^2. \label{Ess}
\end{align}
In this section, we assume that $N_1$ is detectable at colliders, or $M_1|\Theta_{\alpha1}|^2\gg m_j$ for at least one of $\alpha=e,\,\mu,\,\tau$.  The seesaw relation (\ref{Ess}) and the triangle inequality give
\begin{align}
|\Theta_{\alpha1}|^2=\left|\sum_{I=2}^{\cal N}\frac{M_I}{M_1}\Theta_{\alpha I}^2\right|\le\sum_{I=2}^{\cal N}\frac{M_I}{M_1}|\Theta_{\alpha I}|^2\le\frac{\langle \Phi\rangle^2}{M_1M_2}\sum_{I=2}^{\cal N}|F_{\alpha I}|^2.
\end{align}
The third equality holds when $M_2=\dots=M_{\cal N}$. The perturbativity bound can be estimated as 
\begin{align}
|\Theta_{\alpha1}|^2\le \frac{4\pi({\cal N} -1)\langle \Phi\rangle^2}{M_1M_2}, \label{Etha1}
\end{align}
if we apply $|F_{\alpha I}|\le\sqrt{4\pi}$ as the perturbativity conditions. Note that the bound gets more severe when the mass hierarchy is stronger $M_2\gg M_1$. In such a case, the perturbativity conditions on $F_{\alpha I}$ ($I\ge 2$) give the stronger bound on $\Theta_{\alpha 1}$ than the perturbativity of $F_{\alpha 1}$. 
We can use Eq. (\ref{Etha1}) to obtain a lower bound on the mass of the second lightest HNL when $M_1$ and $|\Theta_{\alpha 1}|$ are measured by future experiments. If ${\cal N}_s$ HNLs decouples from the seesaw mechanism (like a dark-matter-candidate HNL), Eq. (\ref{Etha1}) can be slightly modified by ${\cal N}\to {\cal N}-{\cal N}_s$.

We have to take into account the RGE of Yukawa couplings to keep perturbativity at high energy scale. We consider two right-handed neutrinos (${\cal N}=2$) for the calculation of RGE below. 
 For ${\cal N}\ge 3$, some structure of $\Omega$ has to be assumed, but the bound would change only by an ${\cal O}(1)$ factor [see Eq. (\ref{Etha1})]. It is useful to calculate the RGE of neutrino Yukawa couplings in the basis that the active neutrino mass matrix is diagonal, or $y^\nu\equiv -iU^\dag F$. In this basis, our result do not depend on the unknown Dirac and Majorana phases in $U$ because contributions of charged leptons to the RGE are negligible. The mixings $|\Theta_I|^2$ do not depend on the choice of the basis.

\begin{figure}[htb]
\begin{center}
\includegraphics[width=13cm]{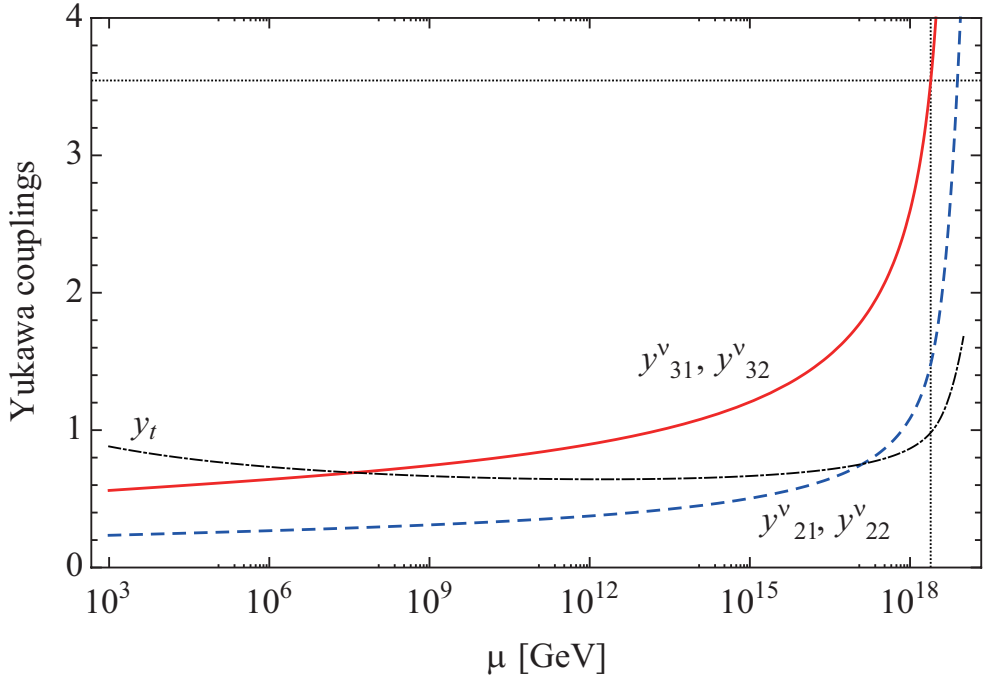}
\caption{\label{Frg} 
Renormalization group evolution of top and neutrino Yukawa couplings $y_t$ and $|y^\nu_{iI}|$ in the case with $M_1=M_2$, $|y_{32}^\nu(M_P)|=\sqrt{4\pi}$ and the NH.}
\end{center}
\end{figure}

A typical example of Yukawa coupling RGE is shown in Fig. \ref{Frg}. We have used one-loop renormalization group equations~\cite{Grzadkowski:1987tf}. The input parameters are taken to the central values in Refs.~\cite{Agashe:2014kda,Xing:2011aa,Gonzalez-Garcia:2014bfa}. In our basis of Eq. (\ref{eq:casas}), $y^\nu_{1I}=0$ for the NH. The absolute values of the other neutrino Yukawa couplings are monotonically increasing, so we define the perturbativity condition here as
\begin{align}
|y_{jI}^\nu(\mu=M_P)|< \sqrt{4\pi}
\end{align}
where $\mu$ is a renormalization scale and $M_P=2.4\times 10^{18}$ GeV (the difference of the results from those applying $|F_{\alpha I}(\mu=M_P)|<\sqrt{4\pi}$ is negligible). The perturbativity of the top Yukawa coupling is satisfied since it is smaller than the maximal component of $|y^\nu|$ at high energy scales.

\begin{figure}[htb]
\begin{center}
\includegraphics[width=11cm]{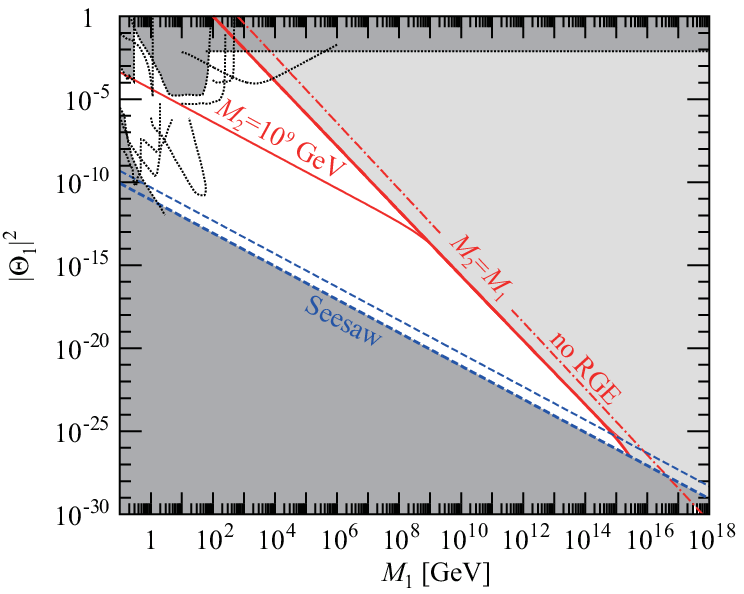}
\vspace{-.3cm}
\caption{\label{Fseesaw} 
 Constraints on the mixing of the lightest HNL. Shaded regions are excluded. The upper bounds from the perturbativity are shown by red solid lines. The bound Eq. (\ref{Etha1}) with ${\cal N}=2,\,M_2=M_1$
is shown by the red dot-dashed line labeled ``no RGE''. Blue dashed lines are the bounds Eq. (\ref{eq:low}) for the NH and IH. Dotted lines show experimental and cosmological bounds and future sensitivities \cite{Deppisch:2015qwa,Antusch:2015mia,Asaka:2015oia}.
}
\end{center}
\end{figure}

Constraints on the mixing $|\Theta_1|^2$ is shown in Fig.~\ref{Fseesaw}. We have drawn the perturbativity bound with $M_2=M_1$ for both the NH and IH cases, but the difference is negligible. The perturbativity bound falls as $\propto M_1^{-2}$ when $M_2=M_1$ and $\propto M_1^{-1}$ when $M_2$ is fixed [see Eq. (\ref{Etha1})]. We can see that two HNLs cannot be heavier than ${\cal O} (10^{15})$~GeV. It should be noted that the seesaw lower bound Eq. (\ref{eq:low}) can be weaker or stronger for ${\cal N} = 3$ depending on the lightest neutrino mass. 

\begin{figure}[htb]
\begin{center}
\includegraphics[width=11cm]{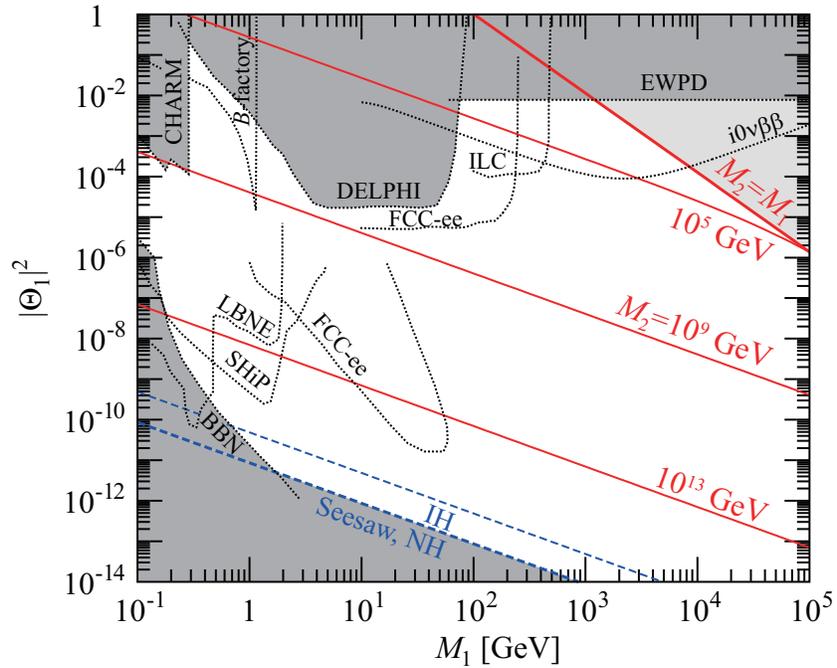}
\vspace{-.3cm}
\caption{\label{Fzoom} 
Low mass region of Fig. \ref{Fseesaw}. The labels of dotted lines indicate names of experiments or methods \cite{Deppisch:2015qwa,Antusch:2015mia,Asaka:2015oia}.
}
\end{center}
\end{figure}

The low mass region is depicted in Fig. \ref{Fzoom} with the constraints from various experiments (see Refs. \cite{Deppisch:2015qwa,Asaka:2015oia} and references therein). The constraints are mostly dominated by those on $\Theta_{\tau 1}$.  The perturbativity bound can be much stronger than those of past experiments, and also comparable to the sensitivities of future experiments.

Note that Fig. \ref{Fzoom} says that low energy experiments can constrain HNLs much heavier than the electroweak scale, as indicated by Eq. (\ref{Etha1}). The constraint is not much changed by $\cal N$ [see Eq. (\ref{Etha1})]. If $M_2>10^9$ GeV, the mixing $|\Theta_1|^2$ have to be smaller than the line labeled ``$M_2=10^9$ GeV''. For example, if a HNL with $M_1\sim10$ GeV and $|\Theta_{1}|^2\sim 10^{-5}$  is found, the second lightest HNL must be lighter than $\sim 10^9$ GeV by the perturbativity and the seesaw relation. It means that thermal leptogenesis \cite{Fukugita:1986hr}, which requires two HNLs heavier than $10^9$ GeV to explain the baryon asymmetry of the universe \cite{Hamaguchi:2001gw,Davidson:2002qv,Giudice:2003jh}, is disfavored for ${\cal N}=3$. In this case, the baryon asymmetry have to be generated by another mechanism, such as resonant leptogenesis \cite{Pilaftsis:1997jf} or baryogenesis via neutrino oscillations  \cite{Akhmedov:1998qx,Asaka:2005pn}.

\section{Conclusion}
We have explicitly shown two constraints on right-handed neutrinos. The lower bound comes from the seesaw relation, and the upper one comes from perturbativity. The mixing of the lightest HNL is related to those of heavier HNLs via the seesaw mechanism. If the lightest HNL is found by experiments, the second lightest HNL have to be sufficiently light. This fact can be used to probe high energy phenomena such as leptogenesis.

\section*{Acknowledgements}
T.A. was supported by JSPS KAKENHI Grants No.
25400249, No. 26105508, and No. 15H01031.

\clearpage


\end{document}